\documentclass[11pt]{article}
\usepackage{graphicx}
\usepackage{amsmath}
\usepackage{amssymb}
\usepackage{amsfonts}
\usepackage{latexsym}
\usepackage{color}
\usepackage{hyperref}
\catcode `\@=11 \@addtoreset{equation}{section}

\catcode `\@=12

\usepackage{color}
\usepackage{graphicx}

\begin{document}
	
	\thispagestyle{empty}

	\vspace*{2cm}
	
	\begin{center}
		{\Large \bf   Integral Transforms and  $\mathcal{PT}$-symmetric Hamiltonians}   \vspace{2cm}\\
		
		{\small  M. W. AlMasri  \footnote{Email: mwalmasri2003@gmail.com}}\\ 
		\small  CyberSecurity and  Systems Research Unit, ISI-USIM,\\ \small Bandar Baru Nilai, 71800 Nilai, Malaysia\\
		
		\vspace{1cm}
		{\small  M. R. B. Wahiddin}\\   \small
		 CyberSecurity and  Systems Research Unit, ISI-USIM \\ \small  Pusat Tamhidi, USIM , Bandar Baru Nilai, 71800 Nilai, Negeri Sembilan. Malaysia
		
	\end{center}
	
	\vspace*{1cm}
	
	\begin{abstract}
		Motivated by the fact that twice the Fourier transform plays the role of parity operator. We systematically study integral transforms in the case of $\mathcal{PT}$-symmetric Hamiltonian. First, we obtain a closed analytical formula for the exponential Fourier transform of a general $\mathcal{PT}$-symmetric Hamiltonian. Using the Segal-Bargmann transform, we investigate the effect of the Fourier transform on the eigenfunctions of the original Hamiltonian. As an immediate application, we comment on the holomorphic representation of non-Hermitian spin chains, in which the Hamiltonian operator is written in
		terms of analytical phase-space coordinates and their partial derivatives	in the Bargmann space rather than matrices in the vector Hilbert space. Finally, we discuss the effect of integral transforms in the study of the Swanson
		Hamiltonian.
	\end{abstract}
	\newpage
	\section{Introduction}
	In \cite{Bender}, Bender and Boettcher proved the reality of energy eigenvalues for  a large class of non-Hermitian Hamiltonians numerically. The common property between these  Hamiltonians is  their invariance  under the action of  both time-reversal $\mathcal{T}$ and parity $\mathcal{P}$ operations, i.e., $[\mathcal{PT},H]=0$. Since then, many studies have been done in this area with potential applications in optics and other areas of physics such as, open quantum systems, supersymmetric quantum mechanics, topological matter, cold atoms, and magnonic waveguides  \cite{JMP,Dorey, Ahmed,Jones,time,Reboiro,review,Mostafa,book, Musslimani,Klaiman,  Longhi1,Longhi,Ruter,Midya,Garmon,Hatano,Znojil1,Almasri1,Ezawa,Ezawa1,Ezawa2,Hang,Berakdar} .  \vskip 5mm
	In quantum mechanics, wave functions can be written  both in the position ($x$-space) and momentum ($p$-space) spaces using the exponential Fourier transform  \cite{Neumann, Lieber,Dahl}, 
	\begin{eqnarray}
		\phi(p)= \frac{1}{\sqrt{2\pi\hbar}}\int_{-\infty}^{\infty} e^{-ipx/\hbar}\; \psi(x) \; dx, 
	\end{eqnarray}
	and its inverse transform
	\begin{eqnarray}
		\psi(x)= \frac{1}{\sqrt{2\pi\hbar}}\int^{\infty}_{-\infty}e^{ipx/\hbar} \; \phi(p)\;  dp. 
	\end{eqnarray}
	The Schrödinger equation in  the momentum space  reads 
	\begin{eqnarray}\label{schrodinger}
		(E-\frac{p^{2}}{2m})\phi(p)= \frac{1}{\sqrt{2\pi\hbar}} \int^{\infty}_{-\infty} \tilde{V}(p-p^\prime)\; \phi(p^{\prime}) \;dp^{\prime}
	\end{eqnarray}\label{key}
	which is an integral equation. In some situations, it is more convenient to work in  the momentum-space representation such as  delta-potential,  i.e., $V= A\; \delta(x)$\cite{Lieber,Dahl,Samar,Samar1}. Furthermore, in one version of non-commutative quantum mechanics, we have the  canonical  commutation relations  defined as  $[p_{i},p_{j}]=0, [x_{i},x_{j}]=i\theta_{ij}, [x_{i},p_{j}]=i\hbar \delta_{ij}$ where $\theta_{ij}$ is an  antisymmetric tensor with dimension (length)$^2$\cite{Rojas}. In this case, the ordinary product of position-dependent  functions is promoted to the Moyal star-product 
	$f\star g({\bf x})= e^{i\theta_{ij}\partial^{(1)}_{i}\partial^{(2)}_{j}} f({\bf x}_{1}) g({\bf x}_{2})|_{{\bf x}_{1}={\bf x}_{2}={\bf x}}$\cite{Szabo,1,2}. Consequently, working  in the momentum-space might be  more efficient  since the momentum-dependent functions are multiplied in the usual way. Noncommutative quantum mechanics appears naturally in the study of charged particle motion under the effect of magnetic field and thus it has deep connection with Landau levels,  anomalous Hall effect  in ferromagnetic metals and  in the study  of magnetic skyrmions \cite{Gamboa,Horvathy,Demetrian,Szabo1,Li,Nagaosa,Bogdanov,epl,Lux}.
	
	\vskip 5mm  In this work, we present a systematic study of $\mathcal{PT}$-symmetric Hamiltonians under the action of  integral transforms such as the Fourier transform.  Moreover, we use the Segal-Bargmann transform to explore the effect of the Fourier transform on eigenfunctions.  Furthermore, the Bargmann representation of spin operators can be used in the study of non-Hermitian spin chains. In this case, spin operators are written in terms of orthonormal phase-space coordinates obtained from the Segal-Bargmann transform of the coordinate-dependent wavefunctions. Finally, we apply the developed formalism to  the case of  Swanson Hamiltonians. 
	\section{Integral Transforms of $\mathcal{PT}$-symmetric Hamiltonians}
	The integral equation problem is to find the solution to:
	\begin{equation}
		h(x)f(x)= g(x)+ \lambda \int_{a}^{b}K(x,y) f(y)dx,
	\end{equation}
	We are given $h(x), g(x), K(x,y)$ and want to determine $f(x)$. The quantity $\lambda$ is a parameter that may be complex in general, and $K(x,y)$ is called the kernel of the integral equation. If $h(x)=0$, we may take $\lambda=-1$ without loss of generality, and we end up with a Fredholm equation of the first kind or an integral transform. \vskip 5mm
	{\bf Definition 2.1.} {\it The integral transform of a function $f(x)$  in the interval $a\leq x\leq b$ denoted by $\mathcal{I}\{f(x)\}=F(k)$ is \cite{Debnath} 
		\begin{eqnarray}\label{definition}
			\mathcal{I}\{f(x)\}=F(k)= \int^{b}_{a}K(x,k)f(x)dx,
		\end{eqnarray}
		where $K(x,k)$ is called the kernel of the transform and $\mathcal{I}$ is the transform operator}. \vskip 2mm
	The integral transform for a function $f({\bf x})$ with several variables is 
	\begin{eqnarray}
		\mathcal{I}\{f({\bf x})\}= F({\bf k})= \int_{D} K({\bf x},{\bf k})f({\bf x})d{\bf x}, 
	\end{eqnarray}
	where ${\bf x}= (x_{1},x_{2},\dots, x_{n})$, ${\bf k}=(k_{1},k_{2},\dots , k_{n})$ and $D\in \mathbb{R}^{n}$. 
	Apparently $\mathcal{I}$ is a linear operator since 
	\begin{eqnarray}
		\mathcal{I}\{\alpha f(x)+\beta g(x)\}= \int_{a}^{b} \{\alpha f(x)+\beta g(x) \} K(x,k)dx\\ = \alpha \mathcal{I}\{f(x)\}+\beta \mathcal{I}\{g(x)\}. 
	\end{eqnarray}
	\vskip 5mm
	Let $f\in L^{1}(\mathbb{R})$,  the Fourier transform of $f(x)$ in $L^{1}(\mathbb{R})$ is a linear transform which has eigenfunctions obeying 
	\begin{eqnarray}
		\mathcal{F}[\psi]= \lambda\psi, 
	\end{eqnarray}
with $\lambda\in \mathbb{C}$. Specifying the Fourier transform between position and momentum spaces, we define the transform as an  integral of the form 
	\begin{eqnarray}\label{integral}
		\mathcal{F}_{x}[f(x)](k)= F(k)= \int^{\infty}_{-\infty} f(x) \;e^{-ikx}dx. 
	\end{eqnarray}
	where the kernel  of the Fourier transform (in our convention) is $e^{-ikx}$. 
	Clearly, $|\int_{-\infty}^{\infty} e^{-ikx} f(x) dx|\leq \int_{-\infty}^{\infty}|f(x)|dx$. This implies that the integral \ref{integral} exist for all $k\in \mathbb{R}$. One interesting feature of the  Fourier transform is  the generalized Parseval relation. Let $f,g \in L^{2}(\mathbb{R})$ , then 
	\begin{eqnarray}
		\langle f|g\rangle= \int_{-\infty}^{\infty} f \overline{g} \;dx=\frac{1}{2\pi} \int_{-\infty}^{\infty} F\overline{G}\;dk=\frac{1}{2\pi}\langle F|G\rangle
	\end{eqnarray}
Furthermore, if we define the convolution $f*g$ of two functions to be the integral 
\begin{eqnarray}
	f*g(x)=\int_{-\infty}^{\infty}f(x-y)g(y)dy.
\end{eqnarray}
Then, the Fourier transform is 
\begin{eqnarray}
	\mathcal{F}[f*g(x)]=\mathcal{F}[f]\mathcal{F}[g].
\end{eqnarray}
In other words, the Fourier transform of a convolution of two functions $f,g$ is the product of
their Fourier transforms. Other useful identities are 
\begin{eqnarray}
	\mathcal{F}[fg]=\mathcal{F}[f]*\mathcal{F}[g],\\
	\mathcal{F}^{-1}[\mathcal{F}(f)\mathcal{F}(g)]=f*g,\\
	\mathcal{F}^{-1}[\mathcal{F}(f)*\mathcal{F}(g)]=fg.
\end{eqnarray} \vskip 3mm
	It is not difficult to realize that Fourier transform can be viewed as a special case of the bilateral Laplace transform with imaginary argument, 
	\begin{eqnarray}
		F(k)= \mathcal{F}_{x}[f(x)]= \mathcal{L}\{f(x)\}|_{s=ik}
	\end{eqnarray}
	where $\mathcal{L}\{f(x)\}= \int_{-\infty}^{\infty}e^{-sx} f(x)dx$. The  definition \ref{definition}  of integral transforms holds for quantum mechanical systems where the arguments of these transforms are in principle functions of canonical variables $F(\hat{q}_{i},\hat{p}_{i})$ satisfying in general the canonical commutation relation  $[\hat{q}_{i},\hat{p}_{j}]=i\hbar \delta_{ij}$ ( in our case, we simply chose these canonical variables to be the position and momentum operators $x$ and $p$ defined in the continuous Hilbert space).   \vskip 5mm
	The Fourier transform is a four-periodic operator and has an intimate connection with the parity operator $\mathcal{P}f(x)=f(-x)$ since 
	\begin{eqnarray}
		\mathcal{F}^{0}=\mathrm{id},\\
		\mathcal{F}^{1}=\mathcal{F},\\
		\mathcal{F}^{2}=\mathcal{P},\\
		\mathcal{F}^{3}=\mathcal{F}^{-1}=\mathcal{F}\circ\mathcal{P}=\mathcal{P}\circ\mathcal{F},\\
		\mathcal{F}^{4}=\mathrm{id},
	\end{eqnarray}
where $\mathrm{id}$ is the identity operator. From the above relations,  we find that applying twice the Fourier transform is identical to the action of the parity operator.
	\vskip 5mm  
	A Hamiltonian  is called pseudo-Hermitian if there exists a Hermitian operator $\eta$ such that \cite{Mostafa}
\begin{eqnarray}\label{cond}
H^{\dagger}=\eta H\eta 
\end{eqnarray}
The operator $\eta$ is sometimes referred to as the metric operator.  
In this case, the Hermitian indefinite inner product $\langle \langle\;\;|\; \; \rangle\rangle_{\eta}$ is defined by \cite{Mostafa}
\begin{eqnarray}
	\langle\langle \psi_{1}|\psi_{2}\rangle\rangle_{\eta}:= \langle\psi_{1}|\eta|\psi_{2}\rangle
\end{eqnarray}
which is invariant under the time translation generated by the Hamiltonian $H$ if and only if $H$ is $\eta$-pseudo Hermitian. Obviously, when $\eta$ is  the identity operator, i.e., $\eta=\mathrm{id}$, the condition \ref{cond} reduces to the Hermitian case. Since the parity operator is Hermitian, it can be used as a metric operator. In this case, the condition \ref{cond} becomes 
\begin{eqnarray}
	H^{\dagger}=\mathcal{P} H\mathcal{P}
\end{eqnarray}
Hence , in terms of the Fourier operator, it becomes 
\begin{eqnarray}
	H^{\dagger}= \mathcal{F}^{2}H\mathcal{F}^{2}
\end{eqnarray}	\vskip 5mm The next step  is to  derive the expression for the Fourier transform of a general $\mathcal{PT}$-symmetric Hamiltonian. The exponential Fourier transform $\mathcal{F}$ of the operator  $\hat{f}(x)$ is   \cite{Bateman,Korner}
	\begin{equation}
		\mathcal{F}_{x}[\hat{f}(x)]= \hat{F}(k)= \int_{-\infty}^{\infty} \hat{f}(x)e^{-ik\cdot x}dx
	\end{equation}
	and the inverse Fourier transform is defined by 
	\begin{eqnarray}
		\mathcal{F}_{k}^{-1}[\hat{F}(k)]= \hat{f(x)}=2\pi\int^{\infty}_{-\infty} \hat{F}(k)e^{ikx}dk
	\end{eqnarray}
	where $x,k$ are operators in the continuous Hilbert space. 
	Under parity (spatial reflection) $p\rightarrow -p$, $x\rightarrow -x$ and under  time-reversal $p\rightarrow -p$, $x\rightarrow x$ and $i\rightarrow -i$,  the  general $\mathcal{PT}$-symmetric Hamiltonian (in units $\hbar=2m=1$ for simplicity) is
	\begin{eqnarray}\label{main}
		H=p^{2}+ V_{{\mathrm Re}}(x)+i V_{{\mathrm Im}}(x)
	\end{eqnarray}
	where the real part of the potential $V_{\mathrm Re}$ is even  and the imaginary part $V_{{\mathrm Im}}$  is odd under parity transformation.  Note that both $V_{\mathrm Re}$,  $V_{{\mathrm Im}}\in \mathbb{R}$. We define their  exponential  Fourier  transforms as
	\begin{eqnarray}\label{1}
		\tilde{V}_{\mathrm Re}(p)= 2 \int^{\infty}_{0} V_{\mathrm Re}(x) \cos (p\cdot x)\; dx, \\ \label{2}
		\tilde{V}_{{\mathrm Im}}(p)=-2i \int^{\infty}_{0}V_{{\mathrm Im}}(x) \sin(p\cdot x)\;  dx. 
	\end{eqnarray}
We plug \ref{1},\ref{2} in the Schrödinger equation 
	\begin{align}\label{hermiticity}
		\tilde{H}\phi(p)= p^{2}\phi(p)+2\pi \int_{-\infty}^{\infty}\tilde{V}(p-p^{\prime}) \phi(p^{\prime}) dp^{\prime}, 
	\end{align}
	where
	\begin{eqnarray}
		\tilde{V}(p)= 2\int_{0}^{\infty}\left(V_{\mathrm{Re}}\left(x\right) \cos\left(x\cdot p\right)+V_{\mathrm{Im}}\left(x\right) \sin\left(x\cdot p\right)\right) dx.
	\end{eqnarray}
	Interestingly, performing the exponential Fourier transform of the Hamiltonian \ref{main} removes the imaginary unit $i$ in front of $V_{\mathrm{Im}}$ and thus the Hamiltonian becomes apparently Hermitian in the momentum-space representation. Therefore,  the non-Hermitian $\mathcal{PT}$-symmetric Hamiltonian  becomes Hermitian after performing the exponential Fourier transform, as seen in \ref{hermiticity}. Due to the unitarity of the Fourier transform, one could easily prove that in the case of  broken $\mathcal{PT}$-symmetry,  complex pairs  appear  in the Hamiltonian written in the momentum space, and thus  the discrete symmetries of a given Hamiltonian are unaffected by unitary transformations as expected.  
	In the above  we have considered Hamiltonian with one-dimensional position spaces. However, the  generalization to higher-dimensional spaces is straightforward. 
	\vskip 5mm 
	
	Next, we need to investigate the effect of the exponential Fourier transform on the orthonormal basis attached to a quantum system . Consider, for example, the Hermite polynomials , 
	\begin{equation} 
		H_{n}(x)= (-)^{n}\; e^{x^{2}}\;(\frac{d}{dx})^{n}(e^{-x^{2}})\end{equation} 
	 which appears in many quantum mechanical problems, such as quantum harmonic oscillator. The  Fourier transform of Hermite polynomials is \cite{Debnath,Wiener}
	\begin{eqnarray}
		\mathcal{F}\{e^{-x^{2}/2}H_{n}(x)\}=  (-i)^{n} e^{-p^{2}/2} H_{n}(p)
	\end{eqnarray}
	It is worthy to mention that the  imaginary coefficient that appears in the front of the second-hand side $(-i)^{n}$  has no effect on the measured energy eigenvalues. To prove this we compute  the Segal-Bargmann transform of $H_{n}(p)$. Let $f(p)$ be a  function in the momentum  Hilbert space $L^{2}(\mathbb{R}^{d})$, we define the Segal-Bargmann transform with respect to $p$ as   \cite{Bargmann,Segal,Askold,Hall,Wahiddin}
	\begin{eqnarray}
		\mathcal{B}(f(z))= \int_{\mathbb{R}^{d}}e^{-(z \cdot z-2\sqrt{2}z\cdot p-p\cdot p)/2} f(p)\;dp. 
	\end{eqnarray} The Segal-Bargmann transform of   Hermite polynomials $H_{n}$ are simply complex monomials of the form $\{\frac{z^{n}}{\sqrt{n!}}\}$. These monomials form an orthonormal basis since  \begin{equation}
		\frac{1}{\pi}\int_{\mathbb{C}} dz \;e^{-|z|^{2}} \; \overline{z}^{n} z^{m} = n!\; \delta_{mn},
	\end{equation}
	where $z$ is arbitrary complex coordinate (physically, it is the phase-space coordinate $z=x+ip$). As we emphasized in \cite{Almasri}, both $\frac{z^{n}}{\sqrt{n!}}$ and $\frac{(-i z)^{n}}{\sqrt{n!}}$ correspond to the same energy eigenvalue for each level characterized by the same quantum number $n$ and the effect of  $(-i)^{n}$ in  front of Hermite polynomials has no  effect on the measured energy eigenvalues.  \vskip 5mm
	{\bf Definition 2.2.} {\it 	The Bargmann space, also known as the Fock-Bargmann or Segal-Bargmann space, denoted by $\mathcal{H}L^{2}\left(\mathbb{C}^{n}, \mu\right)$, is the space of holomorphic functions with Gaussian integral measure $\mu\left(z\right)= ( \pi)^{-n} e^{-|z|^{2}}$ and  $|z|^{2}= |z_{1}|^{2}+\dots+ |z_{n}|^{2}$}. \vskip 5mm Any entire analytic  function $f\left(z\right)$ in this space satisfy a  square-integrability condition of the type \cite{Hall} 
	\begin{equation}\label{condition}
		||f||^{2}: =\langle f| f\rangle_{\mu}= \left(\pi\right)^{-n}\int_{\mathbb{C}^{n}} |f\left(z\right)|^{2} e^{-|z|^{2}}dz < \infty, 
	\end{equation}
	where  $dz$ is the $2n$-dimensional Lebesgue measure on $\mathbb{C}^{n}=\mathbb{R}^{2n}$. \vskip 5mm
	{\it Remark :} The inner-product between  analytic functions $f\left(z\right)$ and $g\left(z\right)$ in the Bargmann space $\mathcal{H}L^{2}\left(\mathbb{C}^{n}, \mu\right)$ is 
	\begin{equation}\label{inner}
		\langle f|g\rangle_{\mu}= \left(\pi\right)^{-n}\int_{\mathbb{C}^{n}}	 \overline{f}\left(z\right) g\left(z\right) e^{-|z|^{2} } dz. 
	\end{equation}
	{\bf Definition 2.3.} {\it The Segal-Bargmann transform is an integral transformation from $L^{2}(\mathbb{R}^{n})$ to $\mathcal{H}L^{2}(\mathbb{C}^{n},\mu)$ defined as }
	\begin{eqnarray}\label{sb}
		\mathcal{B}(f(z))= \int_{\mathbb{R}^{n}} K_{n}(z,x) f(x)d^{n}x\;\;\; for \; z\in \mathbb{C}^{n} \; and\;  f(x)\in \mathbb{R}^{n}
	\end{eqnarray}
	{\it with $K_{n}(z,x)= e^{\left(z\cdot z-2\sqrt{2}z\cdot x-x\cdot x\right)/2}$ the kernel integral.}\vskip 5mm 
	{\bf Theorem 2.1} (Unitarity of the Segal-Bargmann transform). {\it The Segal-Bargmann transform defined by \ref{sb} is a unitary mapping between the Hilbert space $L^{2}(\mathbb{R}^{n})$ and $\mathcal{H}L^{2}(\mathbb{C}^{n},\mu)$. } 
	\vskip 5mm
	In Bargmann space $\mathcal{H}L^{2}(\mathbb{C},\mu)$, the creation and annihilation operators of the harmonic oscillator are $z$ and $\frac{\partial}{\partial z}$ since 
	\begin{eqnarray}
		\left[\frac{\partial }{\partial z},z\right]f(z)= \frac{\partial }{\partial z}(zf(z))-z\frac{\partial f(z)}{\partial z}=1
	\end{eqnarray}
	similar to the canonical commutation relation $[\hat{a},\hat{a}^{\dagger}]=1$. In this representation, the harmonic oscillator Hamiltonian operator  is $H=\hbar \omega \left(z\frac{d}{dz}+\frac{1}{2}\right)$. Applying to energy eigenstates (Fock states), we find 
	\begin{eqnarray}
		H|n\rangle =\hbar \omega \left(z\frac{d}{dz}+\frac{1}{2}\right)\frac{z^{n}}{\sqrt{n!}}=\hbar \omega \left(n+\frac{1}{2}\right)|n\rangle. 
	\end{eqnarray}

	\section{Non-Hermitian Spin Chains} For a given entire analytic function $f_{\alpha,\beta}(z,w)= \frac{z^{\alpha}}{\sqrt{\alpha !}}\frac{w^{\beta}}{\sqrt{\beta!}}$,  we  define the  spin operators in the two-dimensional Bargmann space $\mathcal{H}L^{2}(\mathbb{C}^{2}, \mu)$ using Bargmann  representation of the Jordan-Schwinger  map as  \cite{Almasri}
	\begin{eqnarray}\label{1}
		S_{x}= \frac{\hbar}{2} \left(z \frac{\partial }{\partial w}+ w \frac{\partial}{\partial z}\right), \\\label{2}
		S_{y}= \frac{\hbar}{2i} \left(z \frac{\partial }{\partial w}-w \frac{\partial }{\partial z}\right), \\\label{3}
		S_{z}= \frac{\hbar}{2}\left(z \frac{\partial }{\partial z}- w\frac{\partial}{\partial w}\right). 
	\end{eqnarray}
	These operators belong to the $SU(2)$ Lie algebra and obey the  commutation relations $[S_{i},S_{j}]= i\hbar\; \varepsilon_{ijk }S_{k}$ since the only non-trivial commutators between $z$, $w$ and their partial derivatives  are $\left[\frac{\partial}{\partial z},z\right]=\left[\frac{\partial }{\partial w},w\right]=1$. The total number operator is 
	\begin{equation}
		N=n_{z}+n_{w}, 
	\end{equation}
	where $n_{z}= z\frac{\partial }{\partial z}$ and $n_{w}=w\frac{\partial }{\partial w}$. \vskip 5mm
	The functions $\{f_{\alpha,\beta}\} $ form an orthonormal basis in the two-dimensional Bargmann space since
	\begin{align}\label{ortho}
		\langle f_{\alpha^{\prime},\beta^{\prime}}(z,w)|f_{\alpha,\beta}(z,w)\rangle_{\mu}\\ \nonumber= \frac{1}{\pi^{2}}\int dz  dw \; \exp[- |z|^{2}-|w|^{2}] \overline{f}_{\alpha^{\prime}, \beta^{\prime}}(z,w)f_{\alpha,\beta}(z,w)\\ \nonumber =  \delta_{\alpha^{\prime},\alpha}\delta_{\beta^{\prime},\beta}
	\end{align}
	where $\alpha,\beta \in \mathbb{N}$ is the set of all natural numbers including zero i.e. $\{0,1,2,\dots\}$. \vskip 5mm 
	One advantage of Bargmann  representation of spin operators appears clearly in the study of spin chains. 
	Consider the non-Hermitian $XX$ spin chain Hamiltonian \cite{Korff1},
	\begin{eqnarray}\label{xx}
		\hat{H}_{s}=\frac{1}{2} \sum_{j=1}^{J-1} \left[S_{j}^{x}S^{x}_{j+1}+ S^{y}_{j}S^{y}_{j+1}+ ig\left(S^{z}_{j}-S^{z}_{j+1}\right)\right],
	\end{eqnarray}
	where for small values of the coupling constant  $g$, it becomes spectral equivalence to the Hermitian Hamiltonian\cite{Korff1}.    
	
	\vskip 5mm
	In Bargmann representation, \ref{xx}  can be written analytically as 
	\begin{eqnarray}
		H_{s}=\frac{\hbar^{2}}{4}\sum_{j=1}^{J-1} \left(z_{j+1}w_{j}\frac{\partial^{2}}{\partial z_{j}\partial w_{j+1}}+z_{j} w_{j+1}\frac{\partial^{2}}{\partial w_{j}\partial z_{j+1}}\right)\\ \nonumber
		+ \frac{i\hbar g}{4}\sum_{j=1}^{J-1} \left(z_{j}\frac{\partial }{\partial z_{j}}-z_{j+1}\frac{\partial }{\partial z_{j+1}}+w_{j+1}\frac{\partial}{\partial w_{j+1}}-w_{j}\frac{\partial}{\partial w_{j}}\right).
	\end{eqnarray}
	
	Since the Segal-Bargmann transform is unitary,  the time evolution for spin chain Hamiltonians is the same both in operator and analytical form \cite{Almasri,qda}.
	
	\section{The Swanson Hamiltonian}
	
	The  Swanson Hamiltonian  is \cite{Swanson,Jones1,Fernandez,Bagarello}, 
	\begin{eqnarray}\label{swanson}
		H(\omega,\alpha,\beta) = \hbar \omega\left(\hat{a}^{\dagger}\hat{a}+\frac{1}{2}\right)+\hbar \alpha\; \hat{a}^{2}+\hbar \beta\; \hat{a}^{\dagger\; 2}, 
	\end{eqnarray}
	where $\alpha$ and $\beta$ are real unequal constants in general. When $\alpha=\beta$, \ref{swanson} reduces to the squeezed harmonic oscillator, and the energy spectrum becomes real. \vskip 5mm Eq.\ref{swanson} can be written in terms of position and momentum operators by plugging the exact form of  creation and annihilation operators,  
	\begin{eqnarray}
		\hat{a}=\frac{1}{\sqrt{2}}\left(\frac{x}{\xi_{0}}+i \frac{\xi_{0}}{\hbar} p\right), \\
		\hat{a}^{\dagger}=\frac{1}{\sqrt{2}}\left(\frac{x}{\xi_{0}}-i \frac{\xi_{0}}{\hbar} p\right), 
	\end{eqnarray}
	with $\xi_{0}$ being the characteristic length scale of the system. We obtain 
	
	\begin{eqnarray}\label{swanson1}
		H(\omega,\alpha,\beta)= \frac{1}{2}\hbar \left(\omega+\alpha+\beta\right) \left(\frac{x}{\xi_{0}}\right)^{2}+ \frac{1}{2}\hbar \left(\omega-\alpha-\beta\right)\left(\frac{\xi_{0}p}{\hbar}\right)^{2} +\frac{1}{2}\hbar \left(\alpha-\beta\right) \left(\frac{2i}{\hbar} xp+1\right)
	\end{eqnarray}
	
	It is convenient in our analysis to make the   Swanson-harmonic Hamiltonian mapping. This can be done with proper definitions of the position and momentum operators. Following \cite{Fernandez}, we have 
	\begin{equation}
		P= \left(p+ i\hbar \frac{\alpha-\beta}{(\omega-\alpha-\beta )\xi_{0}^{2}}x\right),
	\end{equation}
	\begin{equation}
		X=x.
	\end{equation}
	It is straightforward to verify the commutation relation $[X,P]=i\hbar$. \vskip 5mm 
	In terms of $X$ and $P$, equation \ref{swanson1} becomes 
	\begin{equation}\label{harmonic}
		H= \frac{1}{2m}P^{2}+ \frac{k}{2} X^{2}\end{equation}
	where 
	\begin{eqnarray}
		m= \frac{\hbar}{(\omega-\alpha-\beta)\xi_{0}^{2}} \in \mathbb{R}, \\
		k= m\Omega^{2}= m \sqrt{\omega^{2}-4\alpha\beta} \in \mathbb{C}. 
	\end{eqnarray}
	As shown by Swanson in \cite{Swanson}, \ref{swanson} posses real and positive energy eigenvalues whenever the condition  $\omega^{2}-4\alpha\beta\geq0$ ($\Omega^{2}\geq 0$)
	is satisfied. Indeed, the limiting case when $\Omega=0$ and $\omega-\alpha-\beta\neq0$ corresponds to a free particle\cite{Fernandez}.
	The creation and annihilation operators are
	\begin{eqnarray}\label{a}
		\hat{A}=\sqrt{ \frac{m\Omega}{2\hbar}}\left(X+\frac{i }{m\Omega} P\right), \\  \label{adagger}
		\hat{A}^{\dagger}=\sqrt{ \frac{m\Omega}{2\hbar}}\left(X-\frac{i }{m\Omega} P\right)
	\end{eqnarray}
	Since $\Omega$ is complex in general, it can be written in the  form $\Omega=|\Omega|e^{i\theta}$. Thus, \ref{a} and \ref{adagger} become
	\begin{eqnarray}
		\hat{A}=\label{A} \sqrt{\frac{m|\Omega|}{2\hbar}}\left(Xe^{i\theta/2}+\frac{i}{m|\Omega|}P e^{-i\theta/2}\right), \\  \label{Adagger}
		\hat{A}^{\dagger}= \sqrt{\frac{m|\Omega|}{2\hbar}}\left(Xe^{-i\theta/2}-\frac{i}{m|\Omega|}P e^{i\theta/2}\right), 
	\end{eqnarray}
	Obviously $[\hat{A},\hat{A}^{\dagger}]=1$. By virtue of Euler's formula, i.e., $e^{i\theta}=\cos\theta+i\sin\theta$, \ref{A} and \ref{Adagger} becomes in matrix form 
	\begin{equation}
		\begin{pmatrix}
			\hat{A}\\
			\hat{A}^{\dagger}
		\end{pmatrix}= \begin{pmatrix}
			\cos\frac{\theta}{2} & i \sin\frac{\theta}{2}\\
			-i\sin\frac{\theta}{2}& \cos\frac{\theta}{2}
		\end{pmatrix} \begin{pmatrix}
			\hat{b}\\
			\hat{b}^{\dagger}
		\end{pmatrix}= {\bf M}\cdot\vec{{\bf b}}
	\end{equation} 
	where $\mathrm{det}({\bf M})= \cos\theta$. When $\theta=2\pi N$ where $N$ is integer, $\mathrm{det}({\bf M})=1$ and this corresponds to unitary transformation, the energy spectrum in this case is real. The operators $\hat{b}$ and $\hat{b}^{\dagger}$ are the usual creation and annihilation operators for a harmonic oscillator with frequency $|\Omega|$ and generalized momentum $P$. Next, consider $|\tilde{n}\rangle$ to be the energy eigenvectors for \ref{harmonic} and $|n\rangle$ to be the energy eigenvectors for a harmonic oscillator with frequency $|\Omega|$, then we have 
	\begin{eqnarray}
		\hat{A}|\tilde{n}\rangle= \sqrt{\tilde{n}}|\tilde{n}-1\rangle=\cos\frac{\theta}{2} \hat{b}|n\rangle+ i\sin\frac{\theta}{2}\hat{b}^{\dagger}|n\rangle\\ \nonumber= \cos\frac{\theta}{2}\sqrt{n}|n-1\rangle+ i \sin\frac{\theta}{2} \sqrt{n+1}|n+1\rangle\\ 
		\hat{A}^{\dagger}|\tilde{n}\rangle= \sqrt{\tilde{n}+1}|\tilde{n}+1\rangle=\cos\frac{\theta}{2} \hat{b}^{\dagger}|n\rangle- i\sin\frac{\theta}{2}\hat{b}|n\rangle\\ \nonumber= \cos\frac{\theta}{2}\sqrt{n+1}|n+1\rangle- i \sin\frac{\theta}{2} \sqrt{n}|n-1\rangle
	\end{eqnarray}
	From these results we notice that when $\sin\frac{\theta}{2}=0$, the Hamiltonian \ref{harmonic} assumes real energy eigenvalues. The potential term in \ref{harmonic} can acquire an imaginary value and thus  break the $\mathcal{PT}$-symmetry. Its exponential Fourier transform has the form 
	\begin{align}\label{final}
		\tilde{V}(P)=\frac{m |\Omega|^{2}}{2} e^{2i\theta} \int_{-\infty}^{\infty}  X^{2} e^{-iP\cdot X} dX
		= -\pi m |\Omega|^{2} e^{2i\theta}\delta^{''}(P)
	\end{align} 
	where $\delta^{\prime\prime}$ is the second-derivative of the Dirac delta function. The non-zero phase $\theta$ breaks the $\mathcal{PT}$-symmetry in the Swanson model. When $\theta=0$, the energy eigenvalues of the Swanson Hamiltonian are real and the $\mathcal{PT}$-symmetry is exact. \vskip 5mm
	The creation and annihilation  operators \ref{A}, \ref{Adagger}  can be written in terms of the operators $\hat{a}$ and $\hat{a}^{\dagger}$ as
	\begin{eqnarray}\label{Aholo}
		\hat{A}= \frac{1}{2} \sqrt{\frac{m|\Omega|}{\hbar}}\left( \left(s\xi_{0}+\frac{\hbar e^{-i\theta/2}}{m|\Omega|\xi_{0}}\right)\hat{a}+ \left(s\xi_{0}-\frac{\hbar e^{-i\theta/2}}{m|\Omega|\xi_{0}}\right)\hat{a}^{\dagger}\right), \\
		\hat{A}^{\dagger}= \frac{1}{2} \sqrt{\frac{m|\Omega|}{\hbar}}\left( \left(s\xi_{0}+\frac{\hbar e^{i\theta/2}}{m|\Omega|\xi_{0}}\right)\hat{a}^{\dagger}+ \left(s\xi_{0}-\frac{\hbar e^{i\theta/2}}{m|\Omega|\xi_{0}}\right)\hat{a}\right)
	\end{eqnarray}
	where $s=\left(e^{i\theta/2}-\frac{\hbar(\alpha-\beta)e^{-i\theta/2}}{m|\Omega|(\omega-\alpha-\beta)\xi_{0}^{2}}\right)$. \vskip 5mm  The operators $\hat{A}$ and $\hat{A}^{\dagger}$ belong to pseudo-boson annihilation and creation operators introduced in \cite{Trifonov}.  Using Bargmann representation, we may write $\hat{A}$ and $\hat{A}^{\dagger}$ respectively, as 
	\begin{eqnarray}
		\frac{\partial }{\partial w}= \frac{1}{2} \sqrt{\frac{m|\Omega|}{\hbar}}\left( \left(s\xi_{0}+\frac{\hbar e^{-i\theta/2}}{m|\Omega|\xi_{0}}\right)\frac{\partial}{\partial z}+ \left(s\xi_{0}-\frac{\hbar e^{-i\theta/2}}{m|\Omega|\xi_{0}}\right)z\right), \\
		w= \frac{1}{2} \sqrt{\frac{m|\Omega|}{\hbar}}\left( \left(s\xi_{0}+\frac{\hbar e^{i\theta/2}}{m|\Omega|\xi_{0}}\right) z+ \left(s\xi_{0}-\frac{\hbar e^{i\theta/2}}{m|\Omega|\xi_{0}}\right)\frac{\partial}{\partial z}\right),
	\end{eqnarray}
	with $[\frac{\partial}{\partial w},w]=1$.  The eigenfunctions of the Swanson model are monomials of the phase-coordinate $w$ i.e. $\frac{w^{n}}{\sqrt{n!}}$. While the eigenfunctions associated with operators $\hat{a}$ and $\hat{a}^{\dagger}$ are $\frac{z^{n}}{\sqrt{n!}}$ which by construction are the ordinary harmonic oscillator energy states.

	\section{Concluding Remarks} We have investigated the behavior of $\mathcal{PT}$-symmetric Hamiltonians under integral transforms. The fact that twice the Fourier transform serves as a parity operator is used in the inner products of pseudo-Hermitian Hamiltonians. We discussed the effect of applying  the exponential Fourier transform to a non-Hermitian $\mathcal{PT}$-symmetric Hamiltonian and investigated the resulting eigenfunctions in the momentum space.   More interestingly, we applied another type of integral transform called the Segal-Bargmann transform to write  the Hamiltonian in terms of the phase-space coordinates  starting from the Hamiltonians in the position space. This was applied to the case of non-Hermitian $XX$ spin chains which is  spectral equivalent to their Hermitian counterparts in the weak coupling regime. As a model example, we studied the   Swanson model in detail from integral transforms point of view. One possible application of the Fourier transforms in the context of $\mathcal{PT}$-symmetric Hamiltonians can be found in the study of double well delta function potentials   \cite{delta}
	\begin{equation}
		V(x)=\left(1+i\Gamma\right)\delta \left(x+a\right)+\left(1-i\Gamma\right)\delta \left(x-a\right)
	\end{equation}
	which has applications in the study of $\mathcal{PT}$-symmetric  Bose-Einstein condensates. 
	\vskip 5mm
	{\bf Acknowledgment.}	We are grateful to Professor H.F. Jones for useful email correspondence. We thank the anonymous reviewer(s) for their comments and suggestions. We acknowledge the support from USIM.  
	
\end{document}